# Preprint

**A Scalable Factory Backbone for Multiple Independent Time-Sensitive Networks**

Wolfram Lautenschlaeger, Florian Frick, Konstantinos Christodoulopoulos, Torben Henke

# A Scalable Factory Backbone for Multiple Independent Time-Sensitive Networks

Wolfram Lautenschlaeger[1], Florian Frick[2], Konstantinos Christodoulopoulos[1], Torben Henke[2]

[1] *Nokia Bell Labs, Stuttgart, Germany*
[2] *ISW, University of Stuttgart, Stuttgart, Germany*

*Abstract*—Convergence of time-sensitive machine control networks as part of the operational technology (OT) with the ubiquitous information technology (IT) networks is an essential requirement for the ongoing digitalization of production. In this paper, we review the fundamental differences between both technologies, the challenges to be solved, existing and upcoming solutions like TSN and their limitations. Furthermore, we introduce an Ethernet extension for a backbone network at factory scale and line rates of 10 – 100Gbit/s. The backbone is intended to carry massive amounts of IT traffic together with the traffic of multiple independent OT networks at the precision of leading-edge field bus technologies in the sub-microsecond range. The backbone remains transparent and does not require changes to the attached OT sub-networks. We prove our claims by prototype measurements, interoperability tests and field trials.

*Keywords*—Time-Sensitive Networking, industrial Ethernet, network convergence, time synchronization, jitter, time division multiplexing

## 1. INTRODUCTION

The networking technology used for industrial communication in factories and production environments evolved into two distinct directions. Information technology (IT) networks are proliferating not only offices and data centers, but also production facilities. On the other hand, specialized network technologies evolved for real-time control of machines and production equipment, the so-called operational technology (OT), mainly in the form of field busses. The call for convergence between IT and OT networks is as old as the differentiation is between them. However, recent innovations intensify the demand. New software paradigms and computing principles like cloud computing, big data analytics, machine learning, or pervasive video are massively pushing into the production facilities. Open and standardized application interfaces (API) enable frequent and lightweight introduction of innovative ideas. The changes are calling for combined IT and OT capabilities in nearly all devices.

Meanwhile, the transmission layers of IT and OT mostly converged at Ethernet principles. The use of IT commodity silicon in OT products considerably reduced the connectivity cost per device. The picture looks different with regards to network operation. Even though physically compatible, it is impossible to run IT and OT networks over a shared network infrastructure in the large scale today. The IEEE 802.1 Time-Sensitive Networking (TSN) Task Group is extending the Ethernet standard by deterministic capabilities to enable transmission and timing guarantees for certain traffic classes, which allows for the coexistence of IT and OT traffic in a controlled environment, a TSN domain. As such, it is a first step towards a universal backbone solution. The IETF Deterministic Networking (DetNet) working group is currently defining a layer 3 umbrella for interoperability at different degrees of determinism, including but not limited to TSN. DetNet's charter excludes modifications to the transport technology. It operates on the tradeoff between scalability versus timing precision of packet multiplexing.

We extend the TSN idea by a further network layer below packet multiplexing, introducing a time-division multiplexing (TDM) shim layer in the Ethernet protocol stack. The proposed solution resolves the conflict between scalability versus timing precision. The additional shim layer carries multiple instances of independent TSN domains. Independence means each involved TSN domain runs under its own control and responsibility, its own protocol (-versions), notion of time, precision, maintenance- and lifecycle, etc. In contrast to the state-of-the-art, where a TSN domain coincides with a physical island network, the backbone allows a TSN domain to stretch out over different locations without the need for extra cabling between the locations. It is obvious that the additional network layer needs an elaborated orchestration and control at layer 3 and above, which is left for future work. Most likely it will be closely aligned with the DetNet recommendation and standards.

Throughout the paper, we investigate why convergence is so hard to reach, what the core problems to be solved are, which alternative solutions are available and where their limits are. Finally, we present our solution in the form of an additional shim layer in the Ethernet protocol stack. We report a Proof-of-Concept implementation together with lab measurements and field trial results.

## 2. TIME SCALES OF OPERATION

IT networks typically connect processor-controlled devices (servers, laptops, mobile devices). Almost all network protocol functions are executed in software by the operating system or by applications. The software runs in hundreds of concurrent processes on few shared processor cores (CPUs). Whenever something happens (e.g. a data packet arrives), a corresponding piece of software is activated on a CPU that is possibly occupied by some other task. Only after the task switch, the event (arriving packet) can be handled accordingly. The time span between the trigger event and its processing is to certain degree random. It depends on many unpredictable circumstances on the host platform. To give a rough number, a Linux box shows ≈50µs timing jitter on its network interfaces. Any software solution, even with timing guarantees, must inherently be less precise than the uncertainties of the underlying operating system. With multiple such jitter sources combined along the path, it stands to reason that the timing precision of IT applications is hardly better than 1ms.

An IT network itself is typically connected by Ethernet switches or routers. With the assumption of gigabit Ethernet (GE) ports, an Ethernet switch contributes roughly 10 – 20µs of packet jitter if moderately loaded. Compared to the 50µs jitter of the attached computers this is rather small and invisible to the software applications. That is why application designers frequently treat the IT network as something fluid that is always there without any sensible timing implication.

OT networks carry machine-to-machine type control traffic. Compared to IT traffic it is low volume but requires much higher precision. Relevant OT field busses are standardized in IEC 61158 series [1] with corresponding profiles in IEC 61784-2 [2]. In this paper we focus on performance parameters like "Time synchronization accuracy" and "Non-time-based synchronization accuracy", the latter indicating *the maximum jitter of the cyclic behavior of any two nodes, using triggering by periodical events over the network for establishing cyclic behavior* (from [2]). With respect to the former, devices use hardware timestamping for time synchronization at a precision of 10ns (nanoseconds) and below. With respect to the latter, packet arrival jitter down to 0.1µs and below is state-of-the-art. Obviously, the required timing precision of OT is by a factor of 1000 tighter when compared to IT. A switched Ethernet network, which is almost insensible to IT applications, would disrupt most OT applications due to its very coarse operation.

The high precision demand is frequently disputed in the IT networking community. And, indeed, there are many OT applications and profiles with sufficiently loose timing requirements that could be fulfilled by carefully dimensioned IT Ethernet networks. However, these discussions are irrelevant, simply because a backbone as part of an infrastructure should serve *any profile*, not only the easy ones. Respective high precision OT products are available and commercially successful since decades. Moreover, due to the large gap of factor 1000, almost any attempt to fine tune or optimize IT towards OT readiness is likely to fail.

## 3. ROOT CAUSE OF PACKET ARRIVAL JITTER

In a switched packet network, the end-to-end propagation time is the result of multiple factors like fiber/wire propagation, processing time, or waiting times. The fiber/wire propagation delay is unavoidable but constant. The processing delay is an engineering topic and can be made constant. What is not constant are waiting times in packet queues, which is the dominating root cause of packet arrival jitter.

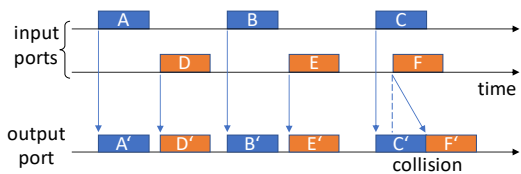

Fig. 1: collision with cross traffic between packets C and F; packet F is unpredictably delayed if compared to D and E of the same stream (assuming cut-through operation)

When two packets arrive almost simultaneously at an output port of a packet switch, only one of them can be served immediately. The other one must wait for the whole forwarding time of the first packet – a packet collision. We use here the term collision for a temporal coincidence, not to be confused with the physical layer collision in the CSMA/CD protocol. On a GE client port, a worst-case packet collision takes 12µs, assuming a 1500-byte packet. Independent packet arrivals on different input ports are an obvious reason for collisions, so-called cross traffic collisions, Fig. 1. However, even with a single input port, collisions are possible, if packets of different size arrive. Due to the store-and-forward principle of operation, small packets are forwarded with shorter delay than large ones. Moreover, they experience a variable waiting time depending on whether a larger packet travels in front of them or not, a self-induced collision, Fig. 2.

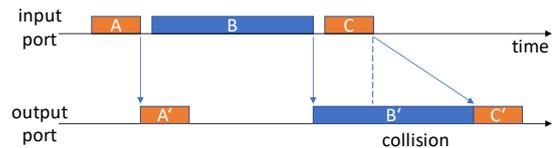

Fig. 2: self-induced collision at store-and-forward; packet B is predictably slower than A due to larger size; equally sized packet C is unpredictably slower than A due to the predecessor B

More than two packets can be involved in a collision, depending on traffic profiles and load. Excess packets build up a queue in a buffer that is subsequently drained, one packet after another. Typical packet buffers in Ethernet switches accommodate 80 – 100 full size Ethernet packets. Accordingly, the unpredictable waiting times (jitter) for queued packets can go up to 100 times the single packet collision duration, roughly 1ms.

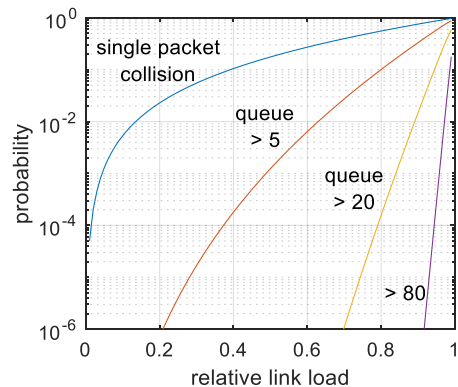

Fig. 3: Probability of waiting times. Large queues fade away at slightly reduced link load < 90%, while single packet collisions remain present even at low load.

Size and probability of large queues can be controlled by appropriate dimensioning of the network capacity. Figure 3 shows the probability for a newly arriving packet to hit (a) another packet (single packet collision), or (b) a queue of certain number of packets, depending on the link load. While with a reduction to 80% link load, the likelihood to encounter a queue of more than 80 packets almost vanishes, the risk of single packet collisions remains high, even at unrealistic low load close to zero.

Admittedly, the discussion above is based on very basic results from queueing theory (M/D/1 queue). It ignores effects like correlated events (packet bursts) or feedback by the TCP protocol. Nevertheless, it illustrates the fundamental difference between queueing in general and its special case packet collision. The impact of queuing is a matter of dimensioning. Whereas packet collision is inherent to packet switching. It creates a jitter floor of at least 12μs (assuming GE ports) almost independent of the traffic load.

## 4. Known Backbone Approaches

A converged factory backbone network in scope of this paper is to be understood as part of the factory infrastructure that provides ubiquitous data connectivity everywhere on the factory site, comparable to electrical power distribution. It is supposed to carry massive amounts of IT traffic together with deterministic high-precision OT traffic. The backbone is not primarily intended for direct attached end-devices. In contrast, it connects technology and application specific island networks to each other, irrespective of their location on the factory site. Given the envisaged amount of IT traffic, a backbone will be based on optical fiber links at l0 – 100Gbit/s line rates.

We further expect backbone distances in the range of one to few kilometers. The resulting round trip delays of at least 10μs preclude most feedback-based approaches like back pressure, flow control, or centralized grant/request scheduling. Similar approaches have been successfully applied to networks on chip (NoC), given the on-chip round trip times in the range of only few nanoseconds [23], [24]. Feedback approaches in IT networking like the TCP protocol, Priority Based Flow Control (PFC) [21], or Software Defined Networking (SDN) [22] provoke waiting times and jitter at least in the range of the round-trip times, which is incompatible to OT traffic.

### 4.1. Traffic engineering

Solutions of this class have in common that they try to control the *queueing jitter* below certain acceptable threshold. Control is done by admission control at the network borders, by traffic shaping, or by statistical considerations. Network calculus [18] plays a prominent role here. Traffic engineering is not intended to and able to avoid *packet collisions*.

The proposals suffer from one weak point: The definition of what is an *acceptable* jitter level. Frequently it is not quantified in absolute values, thus making the impression that it could be controlled at any required low level. But this is not the case due to the remaining floor of single packet collisions. As an example, the Asynchronous Traffic Shaping (ATS) standard [3], [4] provides a formula for the upper delay bound in a mixed traffic scenario. If applied to GE interfaces, it truly predicts the worst-case collision duration of 12μs, independent of the traffic load. Only the queuing delay above that value can be controlled actively. This limitation is not obvious to the reader.

Others argue that the precision requirements of OT applications decrease with increasing distance, e.g. in [19]. This is partially true, because latency matters, too, with increasing distances, which cannot be avoided due to the limited speed of light. On the other hand, the low-precision-at-large-distances paradigm could be a self-fulfilling prophecy. OT applications over large distances avoid strict requirements because the infrastructure does not provide it.

### 4.2. Ethernet extensions

The Ethernet standard IEEE 802.1Q-2018 [21] includes extensions for frame preemption (formerly known as 802.1Qbu) and enhancements for scheduled traffic (802.1Qbv), both frequently subsumed under the task group acronym TSN. Frame preemption enables higher priority frames to interrupt an ongoing transmission of a lower priority frame, thus relaxing the problem of cross traffic collisions, Fig. 1. Without preemption the higher priority frame would have to wait, in the worst-case, for the whole duration of the low priority frame. Preemption solves the collision problem for a single high priority classes against a lower priority class but not within the high priority class itself. To which extent this is acceptable depends on the application.

Scheduled traffic (802.1Qbv) is an extension, where priority queues are served according to a strictly time driven schedule. Gates in front of each priority queue open and close according to a cyclically executed gate control list. If the schedules of all switches in the network are well aligned, time sensitive frames could fly through all the opened gates (no-wait strategy). Other scheduling strategies assume intermediate queuing stops with retiming of the frames by the subsequent gate opening (e.g. cyclic queuing and forwarding in [21]). In any case the time driven scheduling can preserve deterministic frame forwarding for multiple flows in parallel. However, the coordination of the schedules causes a reasonable complexity burden throughout the whole network. For an impression we refer to [20] where Craciunas et al. elaborate the various scheduling constraints that need to be met for safe flow isolation.

### 4.3. Bounded latency and re-timing

This backbone approach uses a setup that permits reasonable, but safely bounded above, arrival jitter. For example, on a 10GE backbone link, that serves only 9 access ports at gigabit Ethernet rate, the queue in front cannot build up larger than 8 maximum sized packets, which corresponds to 8·1518 byte at 10Gbit/s ≈10μs maximum latency variation. At the receiving side, packets can be re-timed with respect to target arrival times. The re-timing requires time stamps associated with the packets and an independent time distribution, which is frequently left to the application layer. The mobile network's eCPRI specification [15] is built on this assumption. Bounded latency and re-timing are popular for high-speed backbones at 100Gbit/s and above, where the queuing delay is low, anyway. However, the rate transition jitter (section 5.2) depends on the client port rate, and as so, it does not benefit from a high-speed backbone.

Many systems use combinations of the aforementioned backbone concepts. The technology of [12] combines a kind of super-preemption over less sensitive IT traffic with a bounded latency approach within the time critical class. The TSN backbone approach of [19] combines strict flow isolation in local TSN domains with bounded latency strategy over the backbone, potentially with re-timing in the receiving TSN domain.

## 5. BACKBONE CHALLENGES

In this section we have a closer look at the reasons, why packet collisions as root cause of jitter (12µs and above) are hard to avoid in a large packet network like a factory backbone.

### 5.1. MULTIPLE TIME DOMAINS

As soon as a factory network exceeds a certain size, it is faced with different time domains. Here the term time domain denotes a set of networked devices that have a common notion of the actual time. In practice, each device runs its own clock that is aligned to some master clock of the domain. While this is obvious at human time scales down to the seconds level, it gets complicated at sub-microsecond precision. Limited signal propagation speed, processing uncertainties, or path asymmetries complicate the alignment of all clocks in a domain. The required time precision depends on the application and ranges from hundreds of milliseconds down to few nanoseconds. The divergent requirements lead to separate time domains per application, vendor, or location, i.e. per machine, production cell, production line, or global, etc. Therefore today, time aware network nodes are required to support the precision time protocol (PTP) with multiple time domains simultaneously (IEEE 1588, IEEE 802.1AS).

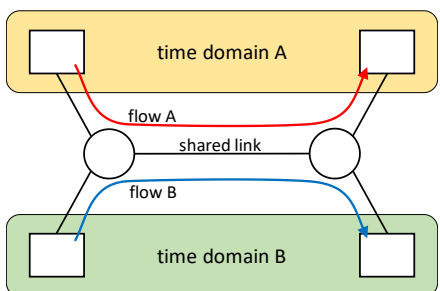

Fig. 4: Different time domains sharing a common packet link

Now, what happens if devices of different time domains communicate over a shared packet network? The case appears, e.g., if production equipment of different vendors in a rough environment shall be connected to their respective detached controllers in a protected control room.

In Fig. 4, a periodic flow A starts and terminates in time domain A, while a similar flow B solely belongs to domain B. Even if both flows adhere to the same nominal repetition rate, their packet arrivals on the shared link cannot be aligned, neither in frequency nor in phase due to the inevitable clock tolerances. Figure 5 illustrates the effect. Initially the red packets arrive well before the blue ones. Over time they move closer and closer and eventually collide with each other. During the collision phase, both flows encounter, one after another, an additional jitter of one packet forwarding delay. After few collisions the packets arrive in opposite order (red is well behind blue). The collision phases re-appear at the beat frequency between the two independent time bases.

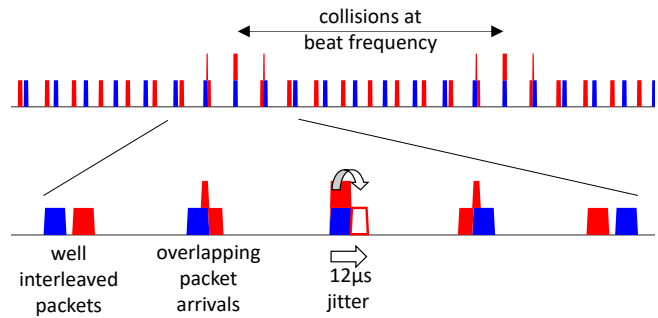

Fig. 5: periodic collisions between packet flows of different time domains; overlapping packet arrivals are served sequentially

The effect can be solved only by explicit synchronization of the two time domains A and B, thus essentially forcing all devices into one unique time domain. It cannot be solved by more precise local oscillators. The collision phase will come eventually. The Ethernet extensions for scheduled traffic (802.1Qbv) would require a combined schedule for the shared link, thus equally forcing all devices into one single time domain.

Interestingly, PTP packets of different time domains experience the same problem of periodic collisions. Time aware switches cannot avoid them. In contrast, they register the extra collision delays for subsequent numeric correction of the measured propagation times (transparent clock mode).

All in all, new PTP revisions solve the problem of multiple time domains *for the PTP itself*, thus enabling *time synchronization accuracy*, section 2. But they cannot prevent the collisions themselves that are impacting the *non-time-based synchronization accuracy* of section 2.

### 5.2. RATE TRANSITION JITTER

High performance IT networks use higher data rates on backbone links compared to the attached client networks and end-devices. E.g. in a network of GE devices, the backbone could be built of 10G Ethernet links. Data center networks with 10GE server interfaces use 40 to 100G backbone links.

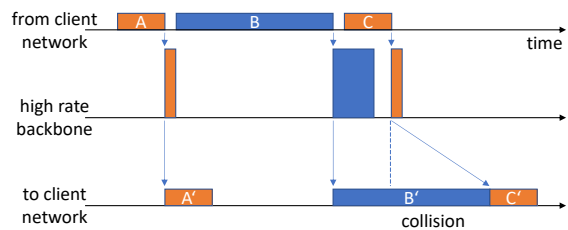

Fig. 6: self-induced collision at rate transition; the position shift at up-conversion causes a collision at the subsequent down-conversion, irrespective of the backbone rate

The rate transition between different link capacities changes the forwarding duration of a packet, given the size in bits remains unchanged. For up-conversion this means, store-and-forward operation is mandatory. Cut-through does not work because the transmission on the faster link can start only after the last bit arrived on the slower link. There is no way to start earlier since the packet length is unknown in advance. In a mixed size packet stream, the combination of up- and down-conversion cause self-induced collisions, similar as of Fig. 2. In a backbone traversal, see Fig. 6, smaller packets encounter different waiting times, depending on whether a larger packet travels in front of them or not. The mutual position between the packets is hard to predict since it changes with every intermediate hop.

## 6. TIME DIVISION MULTIPLEXING

Both backbone challenges, the periodic collisions between different time domains, as well as self-induced collisions at rate transition cannot be avoided in the packet multiplexing layer alone. But they disappear with the introduction of a Time Division Multiplexing (TDM) layer below the packet layer. TDM is the elder digital multiplexing technology compared to packet multiplexing, and it is commonly regarded as cumbersome due to its circuit switching approach. Nevertheless, it is dominating in large operator networks by the Optical Transport Networks technology (OTN).

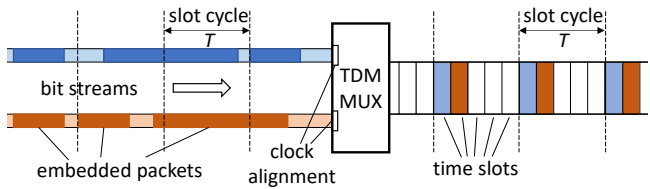

Fig. 7: Time Division Multiplexing of bit streams with embedded packets

TDM systems in a world of packet-based end-devices need three fundamental functions, see Fig. 7: 1. Embedding of packets into a continuous bit stream. 2. Interleaving of bit streams in a periodic schedule of time slots. 3. Clock alignment of asynchronously arriving bit streams. In the following we take a closer look on the TDM functions.

### 6.1. PACKET EMBEDDING INTO A BITSTREAM

Examples of packet embedding are e.g. PoS (Packet over SDH/SONET) [5], [6], where packet boundaries are marked by a control character, while for transparency the occurrence of control characters in the packet body is masked by two byte escape sequences. Masking of control characters implicates some traffic dependent overhead. Packet bodies are expanded by 0.8% in average, in contrast to inter frame gaps that are not expanded.

Another embedding is given by GFP (Generic Framing Procedure) [7], where packet boundaries are indicated by a chain of pointers/length fields that is protected by checksums against false positives in packet data. The length field calculation requires store-and-forward operation with the known jitter impact, Fig. 2. The smallest GFP header is 4 bytes long. So, the original inter frame gaps can be reproduced at a granularity of 4 byte.

The Ethernet physical layer encoding format (10GBASE-R) [8] is an embedding, too, where packets are embedded into a chain of 64B/66B encoded blocks. It uses two-bit block headers to differentiate between data and control blocks. The control blocks in turn encode, among other things, start and termination of a packet. Once again, the packet start position can be encoded at integer 4-byte boundaries.

The established embedding technologies are well approved but not entirely jitter free. They are self-contained, i.e. independent of the enclosing carrier layer. In our architecture proposal, section 8, we introduce a jitter free embedding at the cost of an extra anchor field in the enclosing slot layer.

### 6.2. INTERLEAVING OF BITSTREAMS

Interleaving of bit streams began historically with true pulse (bit) interleaving [9] and grow up with increasing line rates to byte interleaving up to word or block interleaving (e.g. FlexE, [10]) with up to 80 bytes per slot at rates >100Gbit/s. All TDM systems have in common that they use a calendar like schedule that assigns the time slots to client bit streams. The interleaving process implies variable waiting times until the respective slot is due. But, in contrast to packet multiplexing, the waiting time is not entirely random. It depends only on frequency and size of the slot reservations, and not on cross traffic in other slots. Thus, transmit waiting times can be fully compensated by reciprocal waiting times at the reception side.

For estimation of the waiting times it is convenient to imagine the accumulation of one slot volume by client bits in Fig. 7. It takes the time $T=S/b$, where $S$ is the slot size in bit and $b$ is the bit rate of the client bit stream. Since the client bit stream is continuous, a next slot accumulation starts right after completion of the previous one. Hence, the accumulation time $T$ must be identical to the reservation period in the TDM schedule irrespective of the carrier bit rate. In the interplay of accumulation and scheduling, the first bit in a slot must wait at least for the full time $T$, whereas the last bit ideally does not wait at all. For full compensation at the reception side, the first bit is released immediately to the client bit stream, whereas the last bit waits the full period $T$. This way, the total of transmit and receive waiting times is constant and bounded from below by

$$T \geq \frac{S}{b}, \qquad (1)$$

with $S$ the slot size in bit and $b$ the bit rate of the client stream.

Eq. 1 should be kept in mind, when designing a TDM system for low rate client bit streams. TDM transport of single OT command and control streams in the kbit/s range would be either inefficient or cause large delay. For that reason, we are proposing TDM transport of aggregated traffic of whole TSN domains, see section 7, i.e. traffic that, without intermediate TDM layer, and even though scheduled, is vulnerable to self-induced collisions at rate transition, as shown in Fig. 6.

### 6.3. CLOCK ALIGNMENT, BIT STUFFING

Clock alignment is relevant in almost all networking disciplines beyond the chip level. In opposite to networks on chip (NoC) there is usually no extra clock distribution. Clock frequency and phase are implicitly conveyed by the data signal and recovered in the receiver. Due to the limited propagation speed on copper as well as fiber cables (≈ 2/3 speed of light), in a network, the mutual phase differences of arriving signals are unpredictable, even if all nodes would be synchronized to a common grandmaster clock, like e.g. GPS. The phase differences are not even constant due to thermal effects in the cables. Ethernet makes the best of it and entirely forgoes synchronization between nodes. It assumes asynchronous arrival of all physical layer signals. With some restrictions, similar considerations hold for most TDM technologies, too.

The third TDM function, clock alignment, considers that client bit streams arrive asynchronously at the multiplexing point. Even though they all adhere to equal nominal clock frequencies of e.g. 1.25GHz, for cost reasons their real frequency can deviate by e.g. ± 100ppm (parts per million) (Ethernet standard [8]). Without further measure, a similar periodic collision as of Fig. 5 would appear between the slots. The idea of clock alignment is, to raise all bit streams to a common synchronized clock rate by artificially inserting some extra bits or bytes (stuffing bits/bytes), just as much as needed. After demultiplexing, the extra bits or bytes can be discarded to reconstruct the original bit stream.

For example, OTN's asynchronous mapping procedure (AMP) maps the client bit stream into a periodic frame structure [11] with extra header and parity information. (Not to confuse with the OTN time slot, which is one byte.) Two stuff bytes of that frame (NJO, PJO) can be alternatively devoted to the client bit stream or, as extra bytes, to the header. Their presence or absence in the client bit stream is controlled by another header field (JC).

Ethernet uses the variable gaps between packets (inter frame gap, IFG) for clock alignment. The default IFG is 12 idle characters (bytes), the minimum is 5. For clock alignment, idle characters can be duplicated or removed as needed. Since idle characters are meaningless, there is no need to exactly reproduce insertions and deletions after demultiplexing, just the nominal frequency is reconstructed.

## 7. TSN OVER A TDM BACKBONE

In this section we present our vision of a backbone architecture that carries traffic of multiple independent TSN domains in parallel.

TDM alone could solve the backbone challenges of section 5. Packets cannot collide anymore because they travel in different time slots. The necessary waiting times add up to a constant value. Rate transition jitter does not appear, if the client interface rate equals the rate of the TDM slot reservation, which essentially enables true cut-through operation. Finally, the asynchronous mapping for clock alignment avoids any need for synchronization between clients. Unfortunately, TDM is inefficient in carrying low bit rate, command and control like, packet flows. Eq. 1 requires certain minimum capacity reservation to get an acceptably low latency. And a statistical multiplexing gain between low loaded slot reservations does not exist. (We assume here static (pre-configured) reservations and leave out of scope on-demand TDM connections due to their inherent admission delay of, as a rule of thumb, at least one round trip delay.)

On the other hand, TSN scheduled traffic (802.1Qbv) could solve the backbone challenges, too, at the cost of exploding complexity. It would require a factory wide synchronization of all devices at a precision of the most stringent requirement of all applications. The global schedule would have to deal with thousands of flows that are still interacting due to collisions at rate transition, a true complexity nightmare. The paper of Vlk et al. [17] provides a good insight into the complexity of TSN scheduling. In contrast to the intrinsic isolation between TDM slot reservations, TSN's flow isolation is not as strict. It requires a complex per flow interplay of gate control and policing in all involved devices, which makes it prone to misbehaved third party devices. Therefore, non-technical factors further complicate the picture, like different vendors of end-devices, different maintenance and upgrade intervals, different protocol versions, etc. In summary, the Ethernet scheduled traffic extensions are indeed well adapted to transport combined OT and IT traffic in scope of single time domains that natively need to be synchronized due to their joined application, but not more.

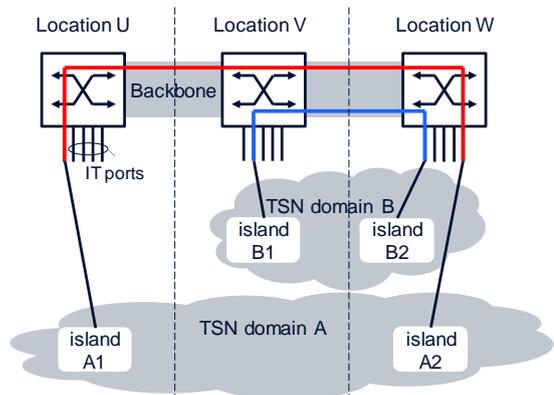

Fig. 8: Backbone vision: Dedicated TDM reservations in the Ethernet PHY for isochronous interconnection of TSN islands; islands A1 and A2 jointly follow their own schedule independent and asynchronous of schedules in domain B or the backbone itself

We propose a TSN over TDM backbone architecture utilized with a TDM shim layer in the Ethernet protocol suite, as shown in Fig. 8. By default, it acts like a best effort (BE) packet switched network, i.e. all time slots are available for best effort IT traffic. On request, the shim layer provides TDM connections between dedicated client Ethernet ports of the switches. The selected ports bypass the packet switching engine and are directly forwarded as a bit stream to the TDM scheduler. Other BE ports remain packet multiplexed, and the combined BE upstream is fed to the TDM layer, where it occupies the remaining unreserved time slots.

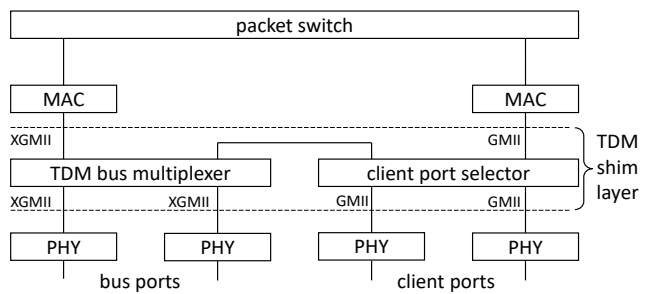

Fig. 9: Layer model of the combined TDM/packet node

From TSN perspective a TDM connection looks like a transparent link, literally a piece of cable. All data and control traffic, scheduled or not, can be transferred over the TDM connection under full control and responsibility of the TSN domain. Different TSN domains use different time slots on the same backbone. The connections are transparent in a sense that they carry any Ethernet frame as is, with no regard to any protocol header. The connections are asynchronous with respect to the backbone clock. Any TSN domain can handle its own notion of time and frequency (protocol, precision, clock master). This does not mean that the backbone should not provide time services, like e.g. SyncE, or a PTP boundary clock. But the TSN domains are free to use it and at which precision.

TDM connections are set-up per TSN domain, not per TSN flow. Therefore, the TSN over TDM backbone configuration remains much coarser than it would be required for an equivalent flat TSN backbone with per flow scheduling.

The transparent connection over a backbone network decouples the TSN time domain from the physical location of the devices. In Fig. 8, the TSN domain **A** is present in locations **U** (e.g. a robot in a rough environment) and **W** (e.g. a detached controller in a protected control room). Scheduled traffic flows at sub-microsecond precision between islands **A1** and **A2**, without interference from/to TSN domain **B** or other IT traffic.

A TDM connection can span over multiple intermediate nodes. To avoid repeated re-scheduling, the backbone links are organized as a *slot synchronous bus* over multiple nodes. Slots of arriving connections are dropped to their respective client ports, new connections are interleaved into the bypassing bus schedule, whereas transiting connections are not processed at all, the slots are just forwarded with the bus. Unreserved slots (BE) are dropped to and re-inserted from the packet switch in every node.

The processing-free node transition minimizes latency and jitter insertion even by long chains of intermediate nodes. But it restricts the architecture to a *linear bus*. Slot synchronous mesh networks are possible on chip [23] but typically fail at larger distances due to mismatching propagation delays. Other, bus-based topologies like fishbone or ladder structures are possible by cross connection nodes (XC) that combine two busses in a node and arbitrate client connections between both. Most likely XC connections will experience latency and jitter implication of two full bus traverses, instead of only one further node transit. (For the difference between both, see Fig. 16)

## 8. AN ETHERNET TDM SHIM LAYER

In this section we draft a shim layer in the Ethernet layer stack that implements the TSN over TDM vision of section 7.

According to the systematic of IEEE 802.3, the new layer is inserted between the MAC and the PCS/PHY layers, both connected by XGMII interfaces. In the layer model of Fig.9, a client port selector decides, which client ports are TDM circuit switched, and which ones are forwarded to the packet switch. A TDM bus multiplexor decides, which TDM slots are forwarded to the other bus interface (transit), which are add/drop traffic to selected client ports, and which are forwarded to the packet switch (best effort).

The shim layer implements a slot synchronous bus by (1) embedding client Ethernet frames into a client bit stream, and by (2) scheduling the client bit stream into bus slots. To do so, it further needs to (3) align the client bit rate to the net rate of the bus, and (4) align the gross rate of the bus to the local clocks of traversed nodes.

Figure 10 shows the hierarchy of client Ethernet over TDM over carrier Ethernet with the corresponding rate alignment operations. The slot clock of the bus (TDM) is generated at the respective head-end node (node #1 in Fig. 10). Downstream nodes (#2 and #3) align their slot phase to the slot arrivals from upstream. Slot phases of opposite bus directions are independent of each other. The bus is not bit synchronous. The bit clock is re-aligned to the local node clock at every node. The client signals A, B, and C arrive asynchronously to the bus. The slot encapsulation (see below) aligns their bit clock to the bus rate. Please note that the client stream A-A' does not undergo individual rate alignment in the transit node #2. Only the bus is rate aligned to the local bit clock.

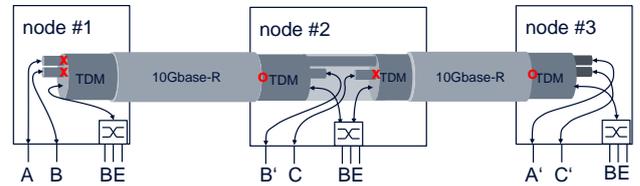

Fig. 10: Embedding hierarchy of client Ethernet over TDM shim layer over carrier Ethernet (10Gbase-R); clock rate alignment (section 6.3): **x** – client Ethernet to TDM, **o** – TDM to local node clock (backward direction omitted); A-A', B-B', and C-C' are independent isochronous connections

The TDM time slots are formatted as regular Ethernet frames in the 10/40/100GBASE-R encoding. Their net size is constant in the range of 100 – 250 bytes, which is left to a design or standardization decision. Slot frames occupy an integer number of 64 bit blocks, including default sized Inter Frame Gaps (IFG), see Fig. 11. This kind of Ethernet encoding preserves the compatibility to well established Ethernet transport technologies. The inter frame gaps enable clock alignment of the bus relative to the local clocks of involved nodes by idle insertion or removal (shim layer function (4) above). On the bus, the slot frames are forwarded across multiple transit nodes without re-coding.

The slot frame uses an own header instead of the Ethernet MAC header. Among others, the header fields indicate the position in the scheduling calendar (function (2)), i.e. which connection it belongs to (a cycle marker, an index into locally stored calendars, source/destination labels, or other means). There is no need to detail it here. Solutions are available from existing standards, e.g. [10], [11]. In the prototype of section 9, we used simple source/destination node/port labels to save the implementation effort for a dedicated control channel.

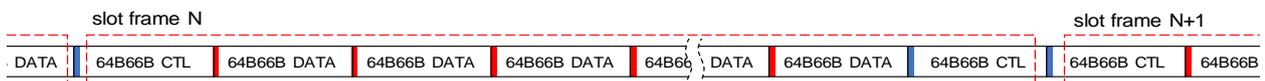

Fig. 11: Slot frame embedded in the sequence of 64-bit blocks of the 10GBASE-R encoding

A justification byte (JB) aligns the client bit stream to the reserved net rate (function (3)). The justification control (JC) flag indicates, if the justification byte belongs to the bit stream or is wasted space in the header, Fig. 12. The resulting justification range of 1 out of 250 bytes equals ±2000ppm, which is big enough to catch any possible combination of Ethernet clock tolerances.

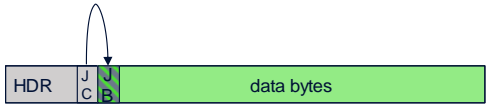

Fig. 12: Rate adaptation by justification control (JC) flag; it assigns the justification byte (JB) to data bytes or not

Embedding of client Ethernet frames into a bit stream (function (1)) could be done by just preserving the GE 8B/10B transport encoding. However, this creates undesired 25% overhead. We chose a pointer-based client frame alignment: An anchor pointer in the slot header points to a first IFG between client frames, where a next pointer points to the next IFG etc. till the end of the slot. Last in the chain is a NULL pointer, as shown in Fig. 13. The pointer occupies only one byte, since the maximum pointer offset is one slot, < 250 byte. Larger IFGs (which is the rule) are filled with pointers just pointing to the next byte. Please note that pointers across slot boundaries are not used. This keeps the pointer small, IFGs are reproduced at byte precision, in contrast to GFP with 4-byte chunks, and the encapsulation time remains constant irrespective of the client frame size. Compared to PoS, section 6.1, it does not expand the payload, but it adds a constant store-and-forward delay of one slot frame for calculation of the anchor pointer. In the prototype we benchmarked both mappings and show the related results in Fig. 15.

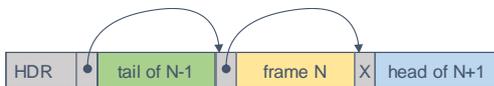

Fig. 13: Pointer alignment of client frames in a slot frame

The unreserved BE slots use the same encapsulation format, however, it is not a fixed rate bitstream, since the BE capacity is the reminder of what is not reserved. Therefore, the BE slots are directly formed from the output queue of the packet switch, at the currently available rate. If multiple packets are waiting in the queue, they are carried back-to-back with just one pointer byte in between.

Finally, a flag in the slot header indicates if a reserved slot is solely filled with idles. This happens on low loaded command and control connections, where timeliness matters but not utilization. If the flag is set, the whole body is reallocated to the BE capacity to the next hop, irrespective of the actual slot destination. The correct number of missed idles is reconstructed at the reservation's destination, given the slot size and the JC flag. The mechanism enables true statistical multiplexing between the traffic in reserved TDM connections and the BE traffic, without impacting the TDM precision. *Even a 100% allocation to low loaded isochronous connections would leave more than enough space to the BE traffic.* One might argue that fragmentation by frequent small packets could inhibit the reuse of whole slot frames. However, preliminary simulations show that this is not the case at lightly loaded (<30%) command and control links.

## 9. PROOF OF CONCEPT

### 9.1. PROTOTYPE IMPLEMENTATION

For verification of our concept, we reused the Optical Ethernet (OE) proof-of-concept (PoC) prototype of [13]. The Optical Ethernet study was targeting at metropolitan area packet networks at line rates of 1Tbit/s and above. Its special focus was on minimalistic transit processing in intermediate nodes (no packet processing, no Forward Error Correction (FEC) re-coding), while preserving the statistical multiplexing capability of packet networks. The result was a slotted bus of large containers, that were subdivided into 56 sub-containers of 160 byte each, mainly for FEC and multiplexing reasons. It enabled opportunistic BE packet traffic access, but also included the concept of isochronous reservations. The prototype implemented the frame layer at downscaled data rate of 10Gbit/s for verification of the protocol. Client ports were 10G Ethernet, too, irrespective of the actual reservation rate, which required a store-and-forward implementation with the known implication of rate transition jitter of section 5.2.

We extended our FPGA boards by native GE ports and implemented 1Gbit/s cut-through add/drop paths from/to the central schedulers in addition to the existing all-10GE design. For latency reasons we reduced the slot size from 56 to 5 sub-containers of 160 bytes each. Instead of the JB/JC byte stuffing (cf. Fig. 12), we preserved the length field of the OE slot header for clock alignment, which implements a rather coarse yet robust byte stuffing by whole sub-containers.

The reservation jitter due to the discrete slot arrival is compensated by two FIFO (first in first out) buffers, one at entry to the bus, one at exit. At entry (Tx side), the GE client frames are embedded into a sequence of sub-containers including all idle spaces, Fig. 13. When the reserved slot is due, the central scheduler picks at most 5 completed sub-containers for transmission but leaves behind what is not yet completed. If the gross reservation rate is slightly above 1Gbit/s, then the actual slot occupation toggles between 4 and 5 sub-containers, thus fulfilling the rate alignment.

At the exit from the bus (Rx side), the slots are unloaded into another FIFO. The FIFO is drained into the client port at nominal GE rate, which is certainly not exactly the same as the originally received GE client rate at entry to the bus. Without further measure, the FIFO would overflow or drain empty, both causing packet corruption and jitter. To avoid this, we implemented a control loop which keeps the FIFO filling as low as possible, but never empty, by sporadic duplication or discarding of idle bytes – the bit stuffing of section 6.3.

The FIFO control implements a proportional-integral (PI) controller with a setpoint (threshold) for the minimum FIFO filling at 14 bytes. The integral component establishes an idle drop/duplication *rate* where threshold violations occur at a defined low ratio of $2^{-17}$. This way it follows the mutual frequency drift of the involved local oscillators. The proportional component triggers per threshold violation one extra idle insertion (or prevents one idle drop, respectively). Both P and I components are required. The proportional part alone exhibits a well-known steady-state error that in the present case translates the frequency drift of the local oscillators into a latency drift. The integral part alone would be unstable because the FIFO as the controlled object is integral, too. And two integrators in a loop are known to oscillate in practice.

The filling of the transmit and receive FIFOs performs a confusing overlay of multiple saw-tooth cycles. Nevertheless, the combined waiting time is constant at a level according to Eq. 1 with a residual jitter in the nanoseconds range.

We verified our architecture by measurements with an IXIA Ethernet tester, by interoperability tests with TSN switches and TSN end-devices, and in a field trial with OTN transport across the city of Stuttgart area, which are described in the following.

*9.2. MEASUREMENT RESULTS*

For lab measurements (Fig. 14), we used a bus of 5 nodes and ran four bidirectional TDM connections in parallel. The experiment emulates a scenario of a central control room at node 1 that supervises field equipment at nodes 2 – 5 in four different TSN domains. For completeness we loaded the best effort channel by 4 – 6 Gbit/s random traffic that we monitored for loss free frame arrival.

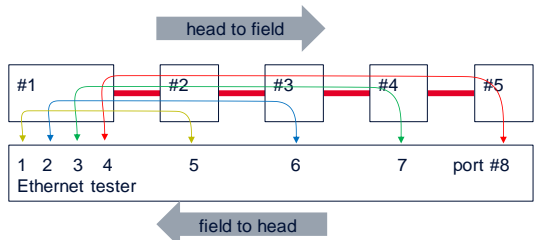

Fig. 14: Backbone scenario emulating a centralized control room at node #1 supervising field equipment at nodes #2 – #5 in different TSN domains

On the shortest connection (#1 – #2) we measured within one hour an average latency of 10µs and a latency variation (peak-to-peak jitter) of 65ns at 800Mbit/s (80% of the client port rate) of random sized packets (64 – 1518 byte). For comparison, a backbone of 10G Ethernet switches with GE client ports results in 16µs latency, including jitter almost of the same amount, measured to be 12µs. Please note that in the prototype experiment all packets from smallest to largest experience the same latency (first bit transmission until first bit arrival). No self-induced collisions and no rate transmission jitter appear along the whole connection.

Figure 15 shows a breakdown of latency contributions: The physical interfaces (Tx+Rx) are standard Ethernet solutions and contribute by 1µs for the GE client port and 0.5µs for the 10GE bus interfaces. The mapping of client frames by IFG pointers (cf. section 8) into the 160-byte sub-containers takes 1.3µs (store-and-forward delay of sub-containers). Alternatively, PoS like mapping by control characters, masking, and scrambling according to RFC1662 [5] takes only 0.3µs at the cost of slightly higher jitter, since the masking of control characters in the frame bodies is traffic dependent. The dominating latency contribution is the compensation of the slot cycle jitter of 6.4µs according to Eq. 1. This term, however, is owed to the larger slot size in the PoC of 5·160=800 byte, if compared to the target architecture, where it should go down to ≈1.3µs at a slot size of e.g. 160 bytes.

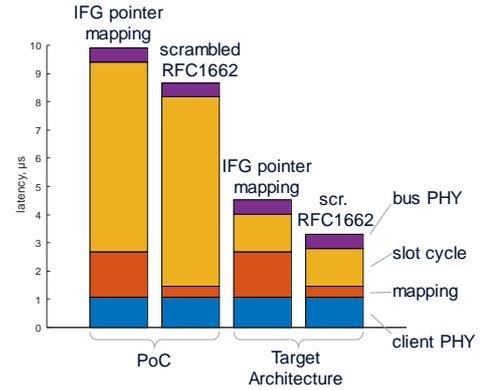

Fig. 15: Measured latency contributions (PoC) and forecast for the target architecture

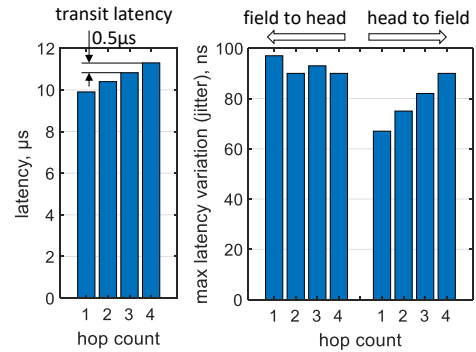

Fig. 16: Impact of hop count along the bus on latency and latency variation within one hour

Figure 16 shows the impact of intermediate transit nodes. The measured latency increases by 0.5µs with every further transit node, which is in line with the bare latency of 10G bus interfaces. Mapping and slot cycle compensation of Eq. 1 do not apply in intermediate transit nodes. The jitter does not show a clear tendency, but at least it is by far less than proportional increase with the number of hops.

*9.3. TSN COMPATIBILITY*

The TSN extensions to Ethernet are not a self-contained approach for a particular purpose, but rather a toolbox for application specific tailored solutions. Therefore, there is no closed form TSN compatibly test. Instead, we prefer a quantitative characterization of the basics that scheduled traffic is relying on.

At first, scheduled traffic requires predictable frame propagation times between TSN devices, otherwise scheduling would not make any sense. We proved this in section 9.2 above, where we demonstrated constant propagation times of mixtures of random sized packets. A TSN application can rely on it, no matter on how its specific schedule is composed.

Second, scheduled devices need a common notion of the actual time. TSN devices use the precision time protocol (PTP, IEEE 1588, IEEE 802.1AS) for time synchronization. PTP communicates by regular Ethernet frames that are equally prone to queueing jitter as any other Ethernet traffic, too. PTP cannot avoid that. Instead it registers the random queuing delays by timestamps and takes them into account in its time calculations. For proper operation, intermediate devices need to report their random delays, too, using a matching protocol revision and profile of PTP, which could be challenging in a heterogeneous production environment.

In our vision, we claim to transport independent TSN domains, potentially including their own synchronization traffic. On the backbone, PTP frames are carried as they are, i.e. protocol agnostic. The backbone's own delay contribution does not need be registered as it is constant like cable propagation.

The following experiment not only evaluates the PTP precision over a TDM connection. It also showcases the correct reproduction of an arbitrary and unpredictable schedule, in this case a schedule of PTP messages and of other background traffic, created by regular Linux PCs.

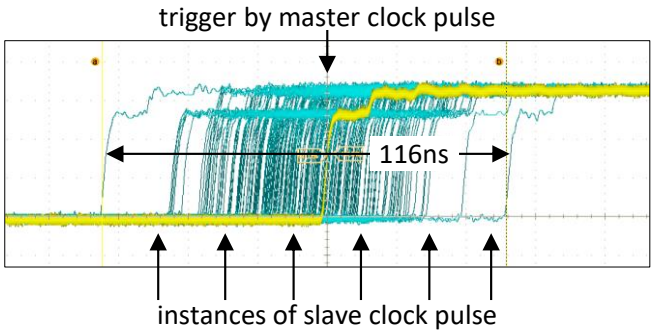

Fig. 17: Precision of PTP clock synchronization across a TDM reservation (oscillogram from [14]), recent results: mean master to slave pulse offset less than 1ns, standard deviation 5.9ns, peak-to-peak 64ns

We used Intel I210 network adapter cards under control of the `linuxptp` driver and watched the electrical 1pps signals (pulse per second) of both cards on a dual channel oscilloscope. In addition to the PTP messages we loaded the same connection/interfaces in both directions at 500Mbit/s random traffic of 1500 byte packets. According to the oscilloscope, see Fig. 17, the mean clock deviation between both cards converged to less than 1ns, whereas the standard deviation of the slave clock pulses, related to the master clock pulses, was 5.9ns. For comparison, the same setup with a direct cable connection resulted in 3.9ns.

The result is remarkable in that our prototype does not implement any PTP feature (neither boundary clock nor transparent clock). With our architecture it is not true anymore that PTP for proper operation requires time awareness of all intermediate nodes. It makes us confident that the architecture can carry any PTP flavor.

*9.4. INDEPENDENCE OF TIME DOMAINS*

With this group of experiments, we assured that the proposed TSN over TDM architecture does not put any additional synchronization burden onto the client networks, neither explicit nor implicit by propagating a backbone clock to the clients. In [14] we reported experiments with four physically independent PTP end-devices that synchronized into two independent master-slave pairs, where the 1pps clock pulses arrived well aligned within each pair but remained floating between the pairs, see Fig. 18.

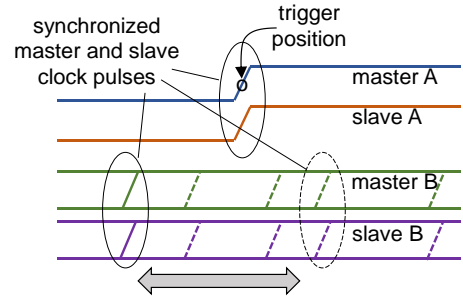

Fig. 18: schematic pulse diagram of two independent PTP time domains; both slave pulses are in sync with the corresponding master pulse; depending on the trigger source (A or B) the respective other pair remains floating

In an extended test we used the Ethernet tester, where we instantiated four PTP instances that we connected over two TDM connections. (Port mapping and traffic offer, except the OTN sections, were chosen identical to the field trial of Fig. 17 and Tab. 1 and 2.) In addition to the PTP control traffic we put 800Mbit/s random sized (64-1518 byte) packet traffic onto each of the TDM connection. For further decorrelation of the flows, we detuned the rate adjustment of each flow individually by small amounts in steps of 0.1%. As expected, the PTP instances arranged into two pairs of grandmaster and slave. The reported maximum difference between arriving PTP synchronization messages and the actual slave clock time stayed in the range of ±40ns.

*9.5. FIELD TRIAL*

In this experiment we proved the compatibility to transport network technologies. The bus signal is a valid 10GBASE-R baseband encoding but without valid MAC frame headers. We connected our prototype boards to an OTN over WDM (Wavelength Division Multiplex) fiber link between the Stuttgart Nokia Bell-Labs premises and the ARENA 2036 research campus, see Fig. 19. The field fiber distance of 25 km was extended by spools of additional 24 km lab fiber. Line termination by Nokia 1830 PSS handled the 10GBASE-R into OTU2e mapping with forward error correction (EFEC2).

For comparison we repeated the lab experiments as above with a loopback connection over two OTN spans. Measurement of the bare OTN loop, 2 x 49km, resulted in 775.23µs latency and <10ns jitter peak-to-peak. The latency of our embedded TDM connections over that loop was 785.54µs with a jitter of 110ns. The slightly larger jitter, compared to Fig. 16, is irrelevant, as it is owed to an earlier prototype release. The two concurrent PTP sessions of section 9.4 converged with a small offset in the range of 10ns due to path asymmetries.

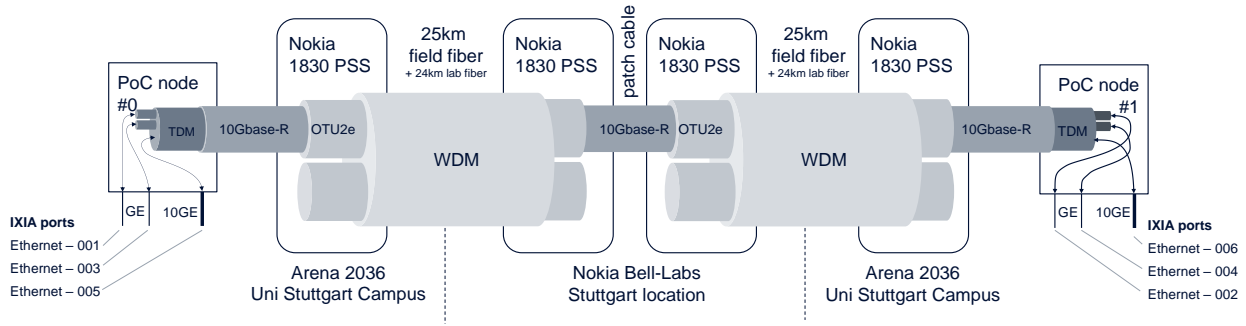

Fig. 19: Protocol stack, topology, and IXIA port mapping of the field trial experiment

TABLE 1: FIELD TRIAL TRAFFIC LOAD AND THE CORRESPONDING LATENCY AND JITTER RESULTS

| Rx Port | Tx Frames | Rx Frames | Frames Delta | Loss % | Rx L1 Rate (bps) | Cut-Through Avg Latency (ns) | Cut-Through Min Latency (ns) | Cut-Through Max Latency (ns) | jitter peak-to-peak (ns) |
|---|---|---|---|---|---|---|---|---|---|
| Ethernet - 002 | 21.946.279 | 21.946.181 | 98 | 0 | 800.842.760 | 785.550 | 785.495 | 785.607 | 112 |
| Ethernet - 001 | 21.946.279 | 21.946.181 | 98 | 0 | 799.479.172 | 785.526 | 785.467 | 785.582 | 115 |
| Ethernet - 004 | 21.946.280 | 21.946.180 | 100 | 0 | 799.503.472 | 785.547 | 785.497 | 785.602 | 105 |
| Ethernet - 003 | 21.946.279 | 21.946.182 | 97 | 0 | 799.485.712 | 785.542 | 785.487 | 785.595 | 108 |
| Ethernet - 006 | 164.597.174 | 164.596.450 | 724 | 0 | 5.999.458.692 | 779.637 | 778.260 | 789.307 | 11.047 |
| Ethernet - 005 | 164.597.174 | 164.596.451 | 723 | 0 | 6.000.543.584 | 779.315 | 777.702 | 789.255 | 11.553 |

TABLE 2: PTP PERFORMANCE RESULTS: TWO INDEPENDENT MASTER-SLAVE SESSIONS IN PARALLEL TO THE TRAFFIC OF TABLE 1

| Port | Protocol | Status | Configured Role | PTP State | Offset [ns] | Max Offset [ns] | Min Offset [ns] | Avg Offset [ns] | Path Delay [ns] | Max Path Delay [ns] | Min Path Delay [ns] | Avg Path Delay [ns] |
|---|---|---|---|---|---|---|---|---|---|---|---|---|
| Ethernet - 001 | PTP 1 | Up | Master | Grandmaster | 0 | 0 | 0 | 0 | 785.550 | 785.590 | 785.530 | 785.551 |
| Ethernet - 002 | PTP 2 | Up | Slave | Slave | -20 | 100 | -60 | 0 | 785.550 | 785.580 | 785.530 | 785.553 |
| Ethernet - 003 | PTP 3 | Up | Master | Grandmaster | 0 | 0 | 0 | 0 | 785.560 | 785.590 | 785.530 | 785.558 |
| Ethernet - 004 | PTP 4 | Up | Slave | Slave | 20 | 80 | -100 | 0 | 785.560 | 785.580 | 785.530 | 785.558 |

## 10. CONCLUSION

The timing precision requirements of operational network technology (OT) are by a factor of 1000 tighter than those of ubiquitous IT networks. Therefore, network convergence cannot be reached just by an improved (upgraded, strengthened, etc.) IT network. For existing packet switched IT networks, we identified the single packet collision, especially at rate transition from client network to backbone and reverse, as the root cause of varying packet arrival times (jitter). It causes a jitter floor of more than 12µs for Gigabit Ethernet clients, almost independent of the network load.

Another challenge for network convergence in the industrial context is the heterogeneity of time domains (vendor specific, protocol versions, maintenance cycles, etc.). The almost impossible time synchronization of a whole factory at sub-microsecond precision provokes packet collisions at the beat frequency between packet flows of different domains.

We reviewed different converged backbone approaches and showed that they mostly cover only part of the challenges but not all at the same time. To overcome this, we introduced a new time-division multiplexing (TDM) architecture as shim layer in the Ethernet protocol stack. We showed in prototype experiments that it enables timing precision of mixed size packet traffic better than 100ns peak-to-peak simultaneously in multiple independent time domains. We verified our claims in interoperability tests with TSN network devices and in a field trial with an OTN transport network.

The current network architecture is limited to a linear bus structure, which is common on the factory floor. For area coverage, however, we are targeting a fishbone or ladder like hierarchy of busses with cross connection nodes. The inevitable timing implications are still under investigation. Further jitter reduction down to the regular Ethernet interface jitter and latency reduction by a more elaborated interleaving are topics of future studies.

The presented work is part of the DDN initiative (Deterministic Dynamic Networks, [16]) and implements reasonable parts of its vision.